\documentclass[twocolumn,aps,showpacs]{revtex4}
\usepackage{amsmath}
\usepackage{latexsym}
\usepackage{float}
\usepackage{amssymb}
\usepackage{graphicx}
\usepackage{epsfig}

\begin{document}

\title{Delocalization-Localization Transition due to Anharmonicity}

\author{David Hajnal}
\author{Rolf Schilling}
\affiliation{Institut f\"ur Physik, Johannes Gutenberg-Universit\"at Mainz, Staudinger Weg 7,
 D-55099 Mainz, Germany}

\date{\today}

\begin{abstract}
Analytical and numerical calculations for a reduced
Fermi-Pasta-Ulam chain demonstrate that energy localization does
not require more than one conserved quantity. Clear
evidence for the existence of a sharp delocalization-localization transition at a
critical amplitude $A_c$ is given. Approaching $A_c$ from above and below, diverging time scales occur.
Above $A_c$, the energy packet converges towards a discrete
breather. Nevertheless, ballistic energy transportation is present,
demonstrating that its existence does not necessarily imply
delocalization.
\end{abstract}


\maketitle

One of the classical investigations of relaxation dynamics of
macroscopic systems is to determine the time evolution of a
perturbed equilibrium state. If this initial state converges to
equilibrium the system is called mixing, implying ergodicity, and it is
nonmixing, otherwise. An important question is: Does there exist a
sharp ergodicity breaking transition under variation of a physical
control parameter like temperature or strength of perturbation?

Within a mode coupling theory for supercooled liquids such a dynamical glass
transition has been found, but its sharpness
seems to result from the mode coupling approximations
(for reviews see \cite{1,2}).
It is not our purpose to contribute to the theory of glass transition, but to study
the influence of anharmonicity on the relaxational behavior at zero-temperature.
In that case the generic lowest energy state of a
particle system is a crystal. One may
ask a similar question as above: Does an initially localized
energy excitation spread over the complete crystal, or not?
In case of small excitation amplitudes, one can use the
\textit{harmonic} approximation. Then the time evolution
of an initial configuration can be determined, exactly \cite{3}.
For one-dimensional harmonic lattices the results are particularly
simple \cite{4,5}. Independent of the strength and size of the
excitation it always spreads over the full system, and energy transportation is ballistic,
provided that there is no disorder. That \textit{infinite} harmonic crystals are ergodic in general,
has been proven rigorously \cite{3a}.
If the excitation amplitude increases, \textit{anharmonicity} gets
important.

Let us neglect any disorder, but taking anharmonic interactions
into account. Discreteness of the lattice combined with
anharmonicity allows for the existence of localized periodic
vibrations, called discrete breathers (DB). For reviews see Ref. \cite{9}.
Their existence suggests that under certain conditions a
\textit{localized} excitation may converge to a DB, whereby
suppressing complete energy spreading. Indeed, numerical solutions
of the discrete nonlinear Schr\"odinger equation (DNLS) \cite{10}
and references wherein, the Klein Gordon chain (KG) \cite{11} and
the $\beta$-Fermi-Pasta-Ulam chain (FPU) \cite{12,13} demonstrate
generation of DB and their role for slow energy relaxation.
Particularly, the numerical results in Refs.~\cite{10,13} give
evidence that DB generation from an initially localized excitation
requires  an excitation amplitude which is large enough.
This has been supported by analytical studies of DNLS and its single impurity version
\cite{13a}.
Concerning analytical
results a little is known, only \cite{14,15,16}. Recently it was
proven for a rather general DNLS (even including disorder) that
the energy spreads incompletely, provided that the norm which is a
measure of anharmonicity, is large enough \cite{16}. This proof is
based on the existence of \textit{two} conserved quantities, the
energy and the norm.

The main questions which now arise are: Does energy localization
need the existence of more than one conservation law? Is there a
sharp transition between complete and incomplete energy spreading?
If so, what are properties characterizing such a
transition? To explore these questions is our
main motivation. Let us consider e.g. the $\beta$-FPU model. In
case that the energy delocalizes completely one can linearize the
equation of motion at large times leading to the harmonic chain,
which is exactly solvable. If it does not, it may converge to a DB.
Since the amplitude of DB decays exponentially, one may linearize again, however
outside the center of the DB. Idealizing this situation we
consider a reduced FPU-chain with one anharmonic bond, only. The
corresponding \textit{classical} Hamiltonian for $N$ particles
with mass $m$ and open boundary condition is as follows:
\begin{eqnarray}\label{eq1}
H &=& \sum \limits _{n=1}^N \frac {1}{2m} p_n^2 + \frac C 2 \sum
\limits _{{n=1, (n\neq M)}} ^{N-1} (x_{n+1} - x_n -a_n)^2 \nonumber\\
&\phantom{x}&+V(x_{M+1} -x_M)
\end{eqnarray}
where $C>0$ is the elastic constant and $a_n$ the
equilibrium length of the $n$-th bond. $V(q)$ is chosen such
that $V$ has a single minimum at $q_{\textrm{min}}=0$ with $V''(q_{min})=C$. This
rules out localized phonon modes of the linearized equation of
motion. In addition we will assume that $V''(q)$ is
increasing with increasing $|q-q_{\textrm{min}}|$.
$H$ is translationally invariant. After
separation of c.o.m and introducing relative coordinates
$q_n=x_{n+1}-x_n-a_n$ and their conjugate momenta $\pi_n$, the
harmonic part of Eq.~(\ref{eq1}) can be transformed to normal
coordinates $\{Q_\nu^L,P_\nu^L\}$ and $\{Q_\mu^R,P_\mu^R\}$ for
the left $(1 \leq n \leq M-1$) and right $(M+1 \leq n \leq N)$
harmonic part of the chain. Skipping the c.o.m.-energy, this yields:
\begin{equation}\label{eq2}
H=H_{\textrm{harm}} + H_{\textrm{anh}} +H_{\textrm{int}}
\end{equation}
with $H_{\textrm{anh}} = \frac 1 m \pi ^2_M +V(q_M)$ the Hamiltonain for the isolated anharmonic bond,
the harmonic Hamiltonian $H_{harm}$, and $H_{int}$ containing the interaction of the
anharmonic bond with the harmonic degrees of freedom (d.o.f.).
Hamiltonian (\ref{eq2}) is the only
conserved quantity, after separation of c.o.m.  Since the equation
of motion is linear in the harmonic d.o.f. these can
be exactly eliminated. This leads for $N \rightarrow \infty$ and $M=
{\mathcal{O}}(N)$ to the nonlinear integro-differential equation:
\begin{equation}\label{eq3}
\ddot{q}(\tau)+ \frac {1} {2C} V'(q(\tau))-\int \limits _0^\tau d\tau'
k(\tau - \tau ') \frac 1 C V'(q(\tau'))=0
\end{equation}
where the index $M$ has been dropped for convenience. $Q_\nu^L (0)
\equiv 0,\; P_\nu^L(0)\equiv 0$ and $Q_\mu^R (0) \equiv 0,\;
P_\mu^R(0)\equiv 0$ were chosen as initial conditions. $\tau =
\omega_0t$ is a dimensionless time and $\omega _0=2(C/m)^{1/2}$
the upper phonon band edge. The lower edge is at zero, due
to translation invariance. The memory kernel is given by
$k(\tau)=-\dot {k}_1(\tau)$ where $k_1(\tau)=J_1(\tau)/\tau$
with $J_n$ the Bessel function of order $n$. Having determined for
given initial conditions $q(0)$ and $\dot{q}(0) $ a solution
$q(\tau)$ of Eq. (\ref{eq3}) one obtains the harmonic nearest neighbor bond
coordinates $q_n(\tau)$ from
\begin{equation}\label{eq5}
q_n(\tau)= \int \limits ^\tau _0 d \tau ' G_{|M-n|} (\tau - \tau
')\frac 1 C V'(q(\tau '))\quad,\quad n \neq M
\end{equation}
with the Green function
$G_n(\tau)=2nJ_{2n}(\tau)/\tau.$
As initially localized excitation we choose $q(0)=A$ and $\dot
{q}(0)=0$. Use of a ``velocity excitation'' $q(0)=0$, $\dot{q}(0)=B$
will not change our results qualitatively. With this initial
condition in mind the conservation of the total energy implies that
$|q(\tau)| < A$
for all $\tau > 0$.

As well-known, elimination of a macroscopic number of d.o.f.
induces dissipation. The frequency dependent damping constant
$\gamma (\omega) $ follows from:
\begin{equation}\label{eq9}
\gamma (\omega) = \lim_{\varepsilon \rightarrow 0} \frac 1 \omega \Im\left(
\hat{k}(\omega + i \varepsilon)\right)= \left \{ \begin{array}{r@{\quad ,
\quad}l} \sqrt{1-\omega^2} & |\omega| <1 \\ 0 & |\omega |
\geq 1
\end{array} \right.
\end{equation}
with $\omega$ measured in units of $\omega _0$ and $\hat{k}$ the Laplace transform of $k(\tau)$.
This exact result is obvious. For
$|\omega|<1$, i.e. for frequencies within the phonon band, the
corresponding modes will be damped and consequently decay to zero,
whereas all modes with frequency above that band will be
undamped. If the anharmonic bond is isolated, i.e. the integral
term in Eq.~(\ref{eq3}) is absent, $q(\tau)$ will perform periodic
oscillations with frequency $\Omega_0 (A)$, depending on the
amplitude A. Due to $V''(q_{\textrm{min}})=C$ it follows
for $A \rightarrow 0$ that $\Omega_0 (A) \rightarrow 1/\sqrt{2}$
in units of $\omega _0$. This frequency is within the phonon band.
Since we have chosen a ``hard'' potential, i.e. $d \Omega_0(A)/d A
>0$, there will be a \textit{critical amplitude} $A_c^{(0)}$ such that
$\Omega_0(A)$ touches the upper phonon band edge:
\begin{equation}\label{eq11}
\Omega_0 (A_c^{(0)}) =1 \quad.
\end{equation}

Accordingly, one may speculate that for $A <A_c^{(0)} $ the initial
excitation will completely delocalize and will converge to a
breather for $A >A_c^{(0)}$.
In the following we will chose a symmetric potential $V(x)/C= \frac 1 2
x^2 + \frac 1 4 x^4$ for simplicity. $x$ can be scaled such that the prefactor
of the quartic term equals 1/4. In that case it is
\begin{equation}\label{eq12}
\Omega_0(A)= \frac \pi 4 \sqrt{2+A^2}/K(-A^2/(2+A^2))
\end{equation}
with $K(m)$ the complete elliptic integral of first kind.
Then Eq. (\ref{eq11}) yields:
\begin{equation}\label{eq13}
A_c^{(0)} \cong 1.16715 \quad.
\end{equation}

In order to check the validity of our speculation above, we
determine first the so-called limiting equation for the asymptotic
solution $q_{\infty}(\tau) = \lim _{\Delta \rightarrow \infty}
q(\tau + \Delta)$ \cite{18}. The Laplace transform of Eq.~(\ref{eq3}) taking
into account the initial conditions can be solved for the Laplace
transform $\hat{q}(z)$ of $q(\tau)$ as function of
$\widehat{q^3}(z)$. Transforming back to time regime yields:
\begin{equation}\label{eq14}
q(\tau) = AJ_0(\tau) - \int \limits _0^\tau d\tau ' J_1(\tau -
\tau ')q^3(\tau ')\quad ,
\end{equation}
which is equivalent to Eq.~(\ref{eq3}), as can be proven. For the
pure harmonic chain, i.e. neglecting the nonlinear term, we obtain
directly $q_{\textrm{harm}}(\tau) = A J_0(\tau)$, as is well-known. It is
straightforward to derive the limiting equation:
\begin{equation}\label{eq15}
q_{\infty}(\tau)= - \int \limits _{- \infty}^\tau d\tau ' J_1(\tau
- \tau ') q_{\infty}^3(\tau ')\quad .
\end{equation}

Since $q_{\infty}(\tau)$ is an asymptotic solution not possessing a relaxing component, its Fourier transform $\widetilde{q}_{\infty}(\omega)$ can not have an absolutely continuous part $\widetilde{q}_{\infty}^{(c)}(\omega)$. If it would, its contribution $q_{\infty}^{(c)}(\tau)$ to $q_{\infty}(\tau)$ would relax to zero for $\tau\rightarrow\infty$. Excluding a singular continuous
component (which may occur for disordered systems at the mobility edge), $\widetilde{q}_{\infty}(\omega)$ must have a discrete support, i. e. $q_{\infty}(\tau)$ is either constant, periodic or quasiperiodic. If it is quasiperiodic, then there are at least two incommensurate frequencies $\omega_1$ and $\omega_2$. The anharmonicity generates Fourier modes with frequencies $m_1\omega_1+m_2\omega_2$. There exists an infinite number of integer pairs $(m_1,m_2)$ such that 
$m_1\omega_1+m_2\omega_2$
is within the phonon band. Therefore, these modes are damped (cf. Eq. (\ref{eq9})) and converge to zero. Accordingly, consistent with our numerical results below,
Eq. (\ref{eq15}) has two kind of solutions, only: A static one $q_{\infty}^{static}(\tau)\equiv q_{\infty}$ and a periodic one $q_{\infty}^{periodic}(\tau+\tau_0)\equiv q_{\infty}^{periodic}(\tau)$ with $2\pi/\tau_0>1$ in order to avoid an overlap with the phonon frequencies $|\omega|\leq1$.
Substituting $q_{\infty}^{static}(\tau)\equiv q_{\infty}$ into Eq. (\ref{eq15}) yields the single solution $q_{\infty}^{static}(\tau)\equiv 0$.

\begin{figure}
\includegraphics[width=1\columnwidth]{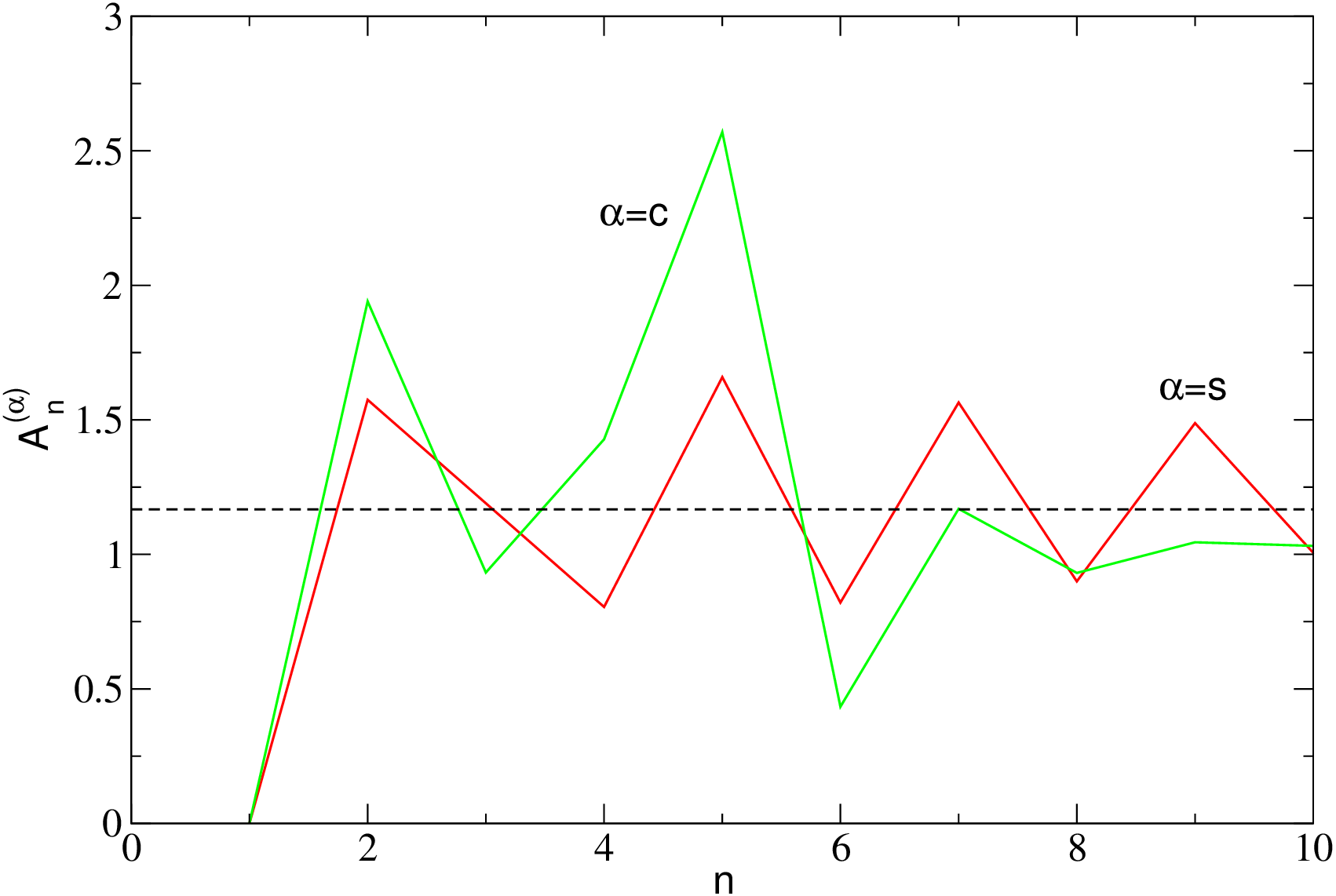}
\caption{\label{fig1}
$n$-dependence of $A_n^{(\alpha)}$ from Eq. (\ref{eq21}) for $\alpha=s,c$.
The dashed line represents $A_c^{(0)} \cong 1.16715$ (cf. Eq. (\ref{eq13})).}
\end{figure}

So far we have argued that two types of asymptotic solutions exist,
a static and a periodic one. In order to
investigate the existence of a critical amplitude $A_c$ we solve
Eq.~(\ref{eq14}) iteratively. With the asymptotic behavior of $J_1$
we arrive at:
\begin{eqnarray}\label{eq20a}
q(\tau) &\cong& A\sqrt{{\frac 2 \pi}} \left[\left(\tau/\tau_s\right)^{-\frac{1}{2}}
\sin \left(\tau - \frac \pi 4\right)\right. \nonumber\\
& &- \left.\left(\tau/\tau_c\right)^{-\frac{1}{2}} \cos
\left(\tau-\frac \pi 4 \right)\right]
\end{eqnarray}
with relaxation times:
\begin{equation}\label{eq20b}
\tau _\alpha (A) = \left[\sum \limits _{n=0}^{\infty} (-1)^n \beta _n
^{(\alpha)} A^{2n}\right]^2\quad, \quad \alpha = s,c\quad .
\end{equation}
$\beta_n^{(\alpha)}$ are given by $n$-fold integrals over products
of $J_1$ and $J_0$. Eq.~(\ref{eq20a}) with $\tau_{\alpha}(A)$ from Eq.~(\ref{eq20b}) is a formal result for
$q(\tau)$ represented by a power series in $A$. It is a physical
solution, only if the infinite sums in Eq. (\ref{eq20b}) do exist.
The critical value $A_c$ is such that this is true for $A<A_c$. Then it is:
\begin{eqnarray}\label{eq21}
A_c &=& min \{A^{(c)}_c,A^{(s)}_c\}\quad,\quad
A^{(\alpha)}_c=\lim_{n\rightarrow \infty}A_n^{(\alpha)}\quad, \nonumber \\
A_n^{(\alpha)}&=&\left|\beta_n^{(\alpha)}/\beta^{(\alpha)}_{n+1}\right|^{1/2}\quad, \quad \alpha = s,c \quad.
\end{eqnarray}
An analytical calculation of these integrals seems impossible. Therefore it is done numerically which leads to $A_n^{(\alpha)}$
shown in Figure \ref{fig1} up to $n=10$. For $n>10$ the numerical errors become significant.
This result gives evidence that $A_c$ is close to $A_c^{(0)}$.
For $A<A_c$ the asymptotic time dependence of $q(\tau)$
is similar to that of the harmonic solution $A
J_0(\tau)$, however with a different phase and a
\textit{renormalized} relaxation time $\tau_{\textrm{rel}}(A) =
\sqrt{\tau _s ^2(A)+\tau _c^2(A)}$, which diverges at $A_c$.
This behavior of $\tau_{rel} (A)$ follows from the divergence of the alternating sums (cf. Eq. (\ref{eq20b}))
due to the quantitative difference of $\beta_n^{(\alpha)}$ for $n$ even and $n$ odd, which also leads to the ``oscillations''
of $A_n^{(\alpha)}$ in Figure \ref{fig1}.
According to Eq. (\ref{eq20a}), $q(\tau)$ decays by an inverse square root
law, as also observed for the original $\beta$-FPU chain \cite{12}.

\begin{figure}
\includegraphics[width=1\columnwidth]{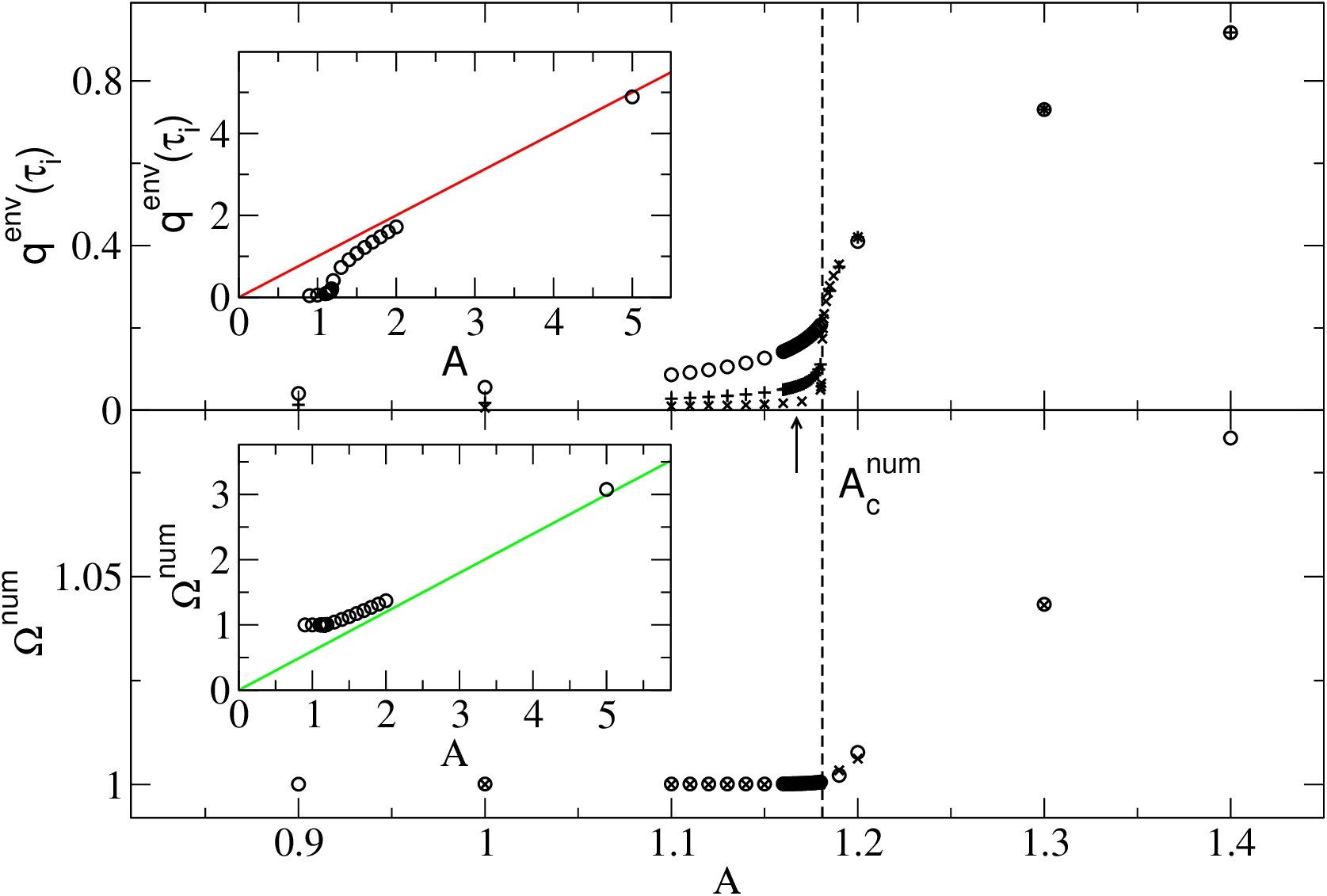}
\caption{\label{fig2}
Top panel: $A$-dependence of $q^{env}(\tau_i)$ for $\tau_i\approx10^3$ (circles), $10^4$ (plus signs) and $10^5$ (crosses) time units. The
inset demonstrates the asymptotic behavior $q^{env}(\tau_i)\sim A$ (solid line).
Bottom panel: DB frequency $\Omega^{num}(A)$ at $\tau_i\approx 10^5$ time units as function of $A$. The arrow indicates the critical
value $A_c^{(0)}$ from Eq. (\ref{eq13}) and the dashed line $A_c^{num}\cong1.181$. The inset shows the asymptotic $A$-dependence
$\Omega(A)\sim A$ (solid line).}
\end{figure}

\begin{figure}
\includegraphics[width=1\columnwidth]{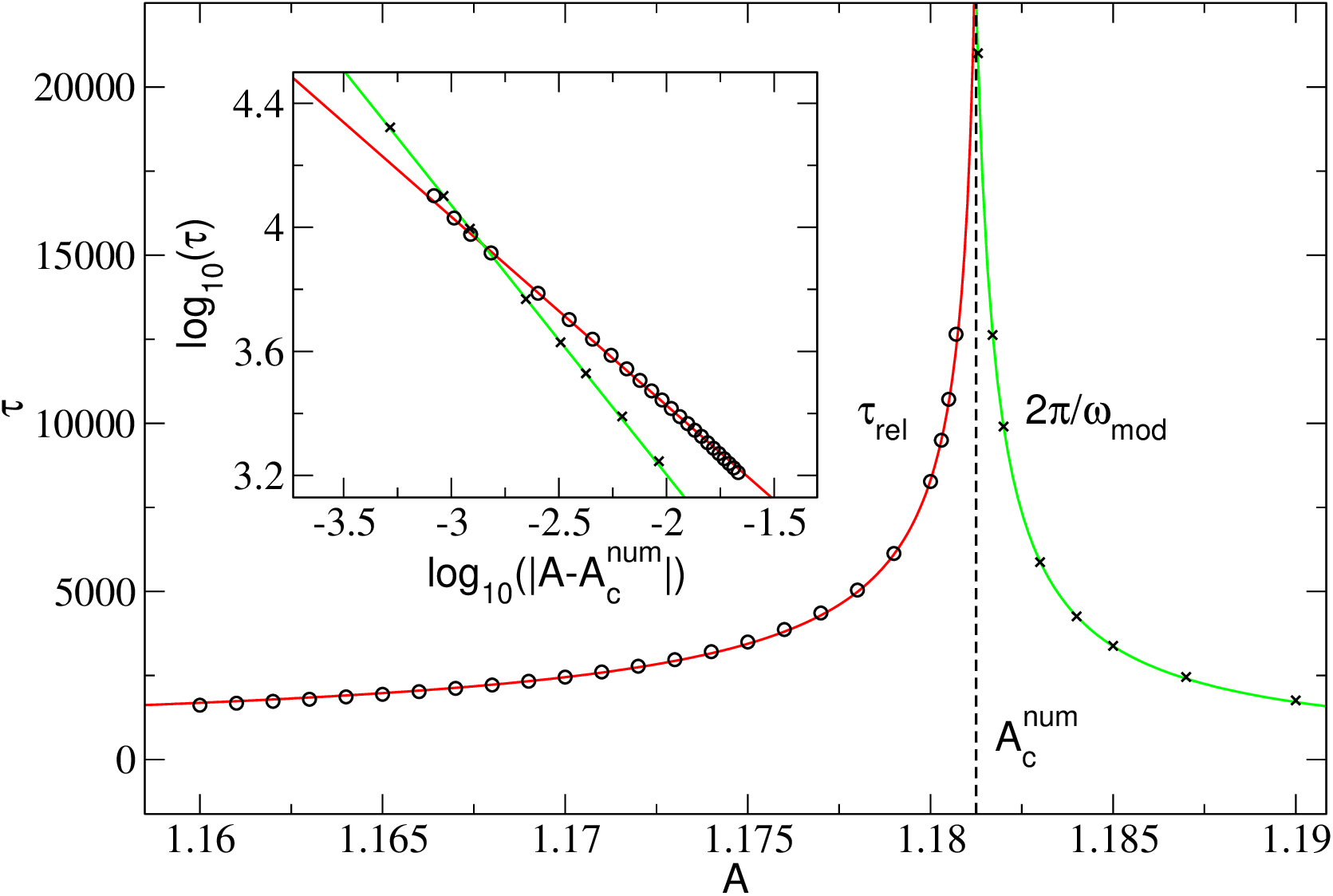}
\caption{\label{fig3}
Renormalized relaxation time $\tau_{rel}$ (circles) and modulation period $2\pi/\omega_{mod}$ (crosses) as function of $A$.
The dashed line indicates $A_c^{num}\cong1.181$. The solid lines represent power law fits of $\tau_{rel}$ and $2\pi/\omega_{mod}$
with exponents $0.61$ and $0.87$, respectively, which are supported by the $\log$-$\log$ plots of the inset.}
\end{figure}

In order to check these results and to access $A >A_c$, we have
solved Eq.~(\ref{eq3}) numerically up to $\tau
_{\textrm{max}}=10^5$ using an integration step of $h=0.05$. Figure ~\ref{fig2} depicts $q^{env}(\tau _i;A)$
for $\tau_i\approx10^3,10^4$ and 10$^5$ where $q^{env}(\tau;A)$ is the envelope function of $|q(\tau)|$ for given $A$.
With increasing $\tau_i$ a clear sharpening of the transition is
found at $A_c^{\textrm{num}} \cong 1.181$, like for a second order
phase transition with finite size effects. $A_c^{\textrm{num}}$ differs from $A_c^{(0)}$ by
about $1.2\%$. The frequency $\Omega^{num}
(A)$ close to $\tau_{\textrm{max}}$ is shown in Figure
\ref{fig2}. For
$A<A_c^{\textrm{num}}$ we have $\Omega^{\textrm{num}}(A) \cong 1$
and for $A>A_c^{num}$ it is well approximated by $\Omega_0(A)$ for the
isolated bond.
However, for $A$ above but close to $A_c^{num}$ the discrepancies are about $2\%$, whereas for $A\gg A_c^{num}$
they disappear. Whether the small deviation of $A_c^{num}$ and $\Omega^{num}$ from $A_c^{(0)}$ and $\Omega_0(A)$,
respectively, is genuine or stems from numerical inaccuracy is unclear. Hence it is not obvious that $A_c=A_c^{(0)}$.
For $A>A_c^{\textrm{num}}$
the initial excitation indeed converges to a DB with frequency
$\Omega^{num} (A)$. Figure \ref{fig3} shows the
numerically determined relaxation time $\tau_{\textrm{rel}}(A)$
for $A<A_c^{num}$, and for $A>A_c^{num}$ the inverse modulation frequency
$2\pi/\omega_{\textrm{mod}} (A)$ of a modulation of the DB, which is
observed numerically. For $\tau \rightarrow \infty$ the modulation
amplitude decays to zero.
$\tau_{rel}$ has been determined from the criterion $q^{env}(\tau_{rel})=A/10$.
Both, $\tau_{\textrm{rel}}$ and
$2\pi/\omega_{mod}$ seem to diverge at $A_c^{num}$  by a power law
with an exponent $\approx0.61$ and $\approx0.87$, respectively
(see inset of Figure \ref{fig3}).
A power law divergence $\tau_{rel}(A)\sim (A_c-A)^{-\gamma}$ implies that $\beta^{(\alpha)}_n\sim(A_c)^{-n}n^{-(1-\gamma/2)}$
for $n\rightarrow\infty$. Whereas the exponential factor is strongly supported by our calculations the validity of the power law part
can not be checked due to the limitation $n\leq 10$.

Finally, we have analytically determined the moments
$m_\ell^{(\textrm{pot})}(\tau)= \sum^N_{n=1} (n-M)^\ell
e_n^{(\textrm{pot})}(\tau)$, $\ell = 1,2,3, \ldots$
of the potential energy profile $e_n^{(\textrm{pot})}(\tau)$ in
the thermodynamic limit. As a result we find
\begin{eqnarray}\label{eq23a}
m_\ell^{(\textrm{pot})} (\tau)&=&\int \limits _0^\tau d\tau _1
\int \limits _0^\tau d \tau _2 K_\ell(\tau - \tau_1,\tau -\tau_2)\nonumber\\
& & \times \frac 1 C V'(q(\tau_1))\frac 1 C V'(q(\tau_2)) \quad.
\end{eqnarray}
Let us restrict to $\ell =2$. $K_2(x,y)$ can be calculated
analytically and expressed by $J_1(x\pm y) $ and $J_2(x\pm y)$.
Taking into account the asymptotic expansion of $J_n$ we find
$m_2^{(\textrm{pot})} (\tau) \sim \tau ^2 \; \textrm{for} \; \tau
\rightarrow \infty,$
for \textit{all} $A$. Hence, the energy transportation is ballistic.
This is expected since the transportation is within the half infinite left and
right \textit{harmonic} part of the chain.
Using the profile of the kinetic energy will not change these
results.

To summarize, based on combined analytical and numerical
calculations of a reduced $\beta$-FPU chain where the
anharmonicity is restricted to a single bond we have presented
clear evidence for the existence of a \textit{critical} amplitude
$A_c$ which separates delocalization from localization. This
demonstrates that a single conservation law is sufficient for
such a transition.
$A_c^{num}$ differs slightly from $A_c^{(0)}$. Therefore it is not clear whether $A_c$
coincides with $A_c^{(0)}$ or not. Of course, no compelling arguments exist for their equality.
In addition, the divergence of the iteration
series is the mathematical origin of the transition at $A_c$ and leads for $A<A_c$ to a renormalized (due to anharmonicity)
relaxation time $\tau_{\textrm{rel}}$ which diverges at $A_c$. The numerical
solution suggests a power law divergence with exponent smaller then one.
Above $A_c$
it yields the convergence towards a DB with frequency very close
to $\Omega_0(A)$ of the isolated bond.
Finally, from the large $\tau$ behavior of the second
moment $m_2(\tau)$ we find \textit{ballistic} energy
transportation for $A<A_c^{num}$ and $A>A_c^{num}$. This proves that a
divergence of $m_2(\tau)$ is not necessarily an indication of
complete energy spreading, as it has been assumed for DNLS
\cite{6}, supporting the conclusion in \cite{16}.

This work was started when one of us (R. S.) was a member of the
Advanced Study Group 2007 at the MPIPKS Dresden. R. S. gratefully
acknowledges the MPIPKS for its hospitality and financial support
and S. Aubry, V. Bach, N. Bl\"umer and S. Flach for stimulating discussions.


\begin{thebibliography}{99}
\bibitem{1} W. G\"otze, ``Complex Dynamics of Glass-Forming Liquids, A Mode-Coupling
Theory'', Oxford University Press, Oxford UK (2008)
\bibitem{2} S. P. Das, Rev. Mod. Phys. \textbf{76}, 785 (2004)
\bibitem{3} W. R. Hamilton, Proc. R. Irish Acad. \textbf{1}, 341
(1841)
\bibitem{4} G. S. Zavt et. al, Phys. Rev.
\textbf{E47}, 4108 (1993)
\bibitem{5} P. K. Datta and K. Kundu, Phys. Rev. \textbf{B51}, 6287 (1995)
\bibitem{3a} J. L. van Hemmen, Lecture Notes \textbf{93}, Springer (1979)
\bibitem{9} S. Flach and C. R. Willis, Phys. Rep. \textbf{295},
181 (1998); S. Aubry, Physica \textbf{D216}, 1 (2006)
\bibitem{10} M. I. Molina and G. P. Tsironis, Physica
\textbf{D65}, 267 (1993); L. J. Bernstein et al., Phys. Lett.
\textbf{A181}, 135 (1993); M. Johansson et al., Phys. Rev.
\textbf{B52}, 231 (1995)
\bibitem{11} G. P. Tsironis and S. Aubry, Phys. Rev. Lett.
\textbf{77}, 5225 (1996)
\bibitem{12} F. Piazza et al, J. Phys. \textbf{A34}, 9803 (2001)
\bibitem{13} R. Reigada et al, Phys. Rev. \textbf{E66}, 046607
(2002)
\bibitem{13a} J. Dorignac, J. Zhou and D. K. Campbell, Physica \textbf{D327}, 486 (2008)
\bibitem{14} S. Flach et al., Phys. Rev. Lett. \textbf{78}, 1207
(1997); M. I. Weinstein, Nonlinearity \textbf{12}, 673 (1999)
\bibitem{15} A. Stefanov and P. G. Kevrekidis, Nonlinearity
\textbf{18}, 1841 (2005)
\bibitem{16} G. Kopidakis et al. Phys. Rev. Lett. \textbf{100},
084103 (2008)
\bibitem{18} R. K. Miller, ``Nonlinear Volterra Integral
Equations'', W. A. Benjamin (1971)
\bibitem{6} A. S. Pikovsky and D. L. Skepelyansky, Phys. Rev.
Lett. \textbf{100}, 094101 (2008)
\end{thebibliography}
\end{document}